\documentclass[fleqn,twoside,twocolumn,nofootinbib]{revtex4} 
\usepackage{ujp} 
\begin{document}
\title[Secondary emission from synthetic opal]
{Secondary emission from synthetic opal infiltrated by colloidal gold and glycine}%
\author{G.I.~Dovbeshko}
\affiliation{Institute of Physics, Nat. Acad. of Sci. of Ukraine}
\address{46, Nauky Av., Kyiv 03028, Ukraine}
\author{O.M.~Fesenko}
\affiliation{Institute of Physics, Nat. Acad. of Sci. of Ukraine}
\address{46, Nauky Av., Kyiv 03028, Ukraine}
\author{V.V.~Boyko}
\affiliation{Institute of Physics, Nat. Acad. of Sci. of Ukraine}
\address{46, Nauky Av., Kyiv 03028, Ukraine}
\email{vb@iop.kiev.ua}
\author{V.R.~Romanyuk}
\affiliation{V.E. Lashkaryov Institute of Semiconductor Physics, Nat. Acad. of Sci. of Ukraine}
\address{41, Nauky Av., Kyiv 03680, Ukraine}
\author{V.S.~Gorelik}
\affiliation{P.N. Lebedev Physical Institute,  Russian Acad. of Sci.}
\address{53, Leninsky Prospect, Moscow 119991, Russia}
\author{V.N.~Moiseyenko}
\affiliation{Dnipropetrovsk National University}
\address{72, Prosp. Gagarina, Dnipropetrovsk 49050, Ukraine}
\author{V.B.~Sobolev}
\affiliation{Technical Center of Nat. Acad. of Sci. of Ukraine}
\address{13, Pokrovskaya Str., Kyiv 04070, Ukraine}
\author{V.V.~Shvalagin}
\affiliation{L.V. Pysarzhevsky Institute of Physical Chemistry, Nat. Acad. of Sci. of Ukraine}
\address{31, Nauky Av., Kyiv 03028, Ukraine}
\udk{533.9} \pacs{42.70.Qs, 68.37.Hk,\\[-3pt] 78.67.Ch} \razd{\secix}

\setcounter{page}{154}%
\maketitle

\begin{abstract}
A comparison of the secondary emission (photoluminescence) and Bragg
reflection spectra of photonic crystals (PC), namely, synthetic
opals, opals infiltrated by colloidal gold, glycine, and a complex
of colloidal gold with glycine is performed. The infiltration of
colloidal gold and a complex of colloidal gold with glycine into the
pores of PC causes a short-wavelength shift (about 5--15 nm) of the
Bragg reflection and increases the intensity of this band by 1.5--3
times. In photoluminescence, the infiltration of PC by colloidal
gold and colloidal gold with glycine suppresses the PC emission band
near 375--450 nm and enhances the shoulder of the stop-zone band of
PC in the region of 470--510 nm. The shape of the observed PC
emission band connected with defects in synthetic opal is determined
by the type of infiltrates and the excitation wavelength. Possible
mechanisms of the effects are discussed.
\end{abstract}

\section{Introduction}

Synthetic opal is known as one of the typical photonic crystals (PC)
\cite{1}. The PC optical property study in the region of the
forbidden zone and its vicinity is of interest for fundamental
investigations and technical applications. A modification of the
stop-zone properties could be done by the infiltration of different
dielectrics, metals, and organic molecules into the pores of PC
\cite{2, 3}. Thus, it was shown in \cite{4} that the minimum of the
luminescence intensity of synthetic opal is registered in a region,
where the Bragg reflection maxima occur. Radiative photon modes go
out freely from PC, while bounded photonic modes always exist inside
the globules of big sizes. These modes are not radiative due to
their total internal reflection on the surface of globule sphere.

The introduction of NaNO$_{2}$ (non-organic dye) into the PC pores
has led to changes of the secondary emission spectra \cite{5, 6}.
Namely, in a thin crystal for the transmittance mode, the
fluorescence of NaNO$_{2}$ is suppressed, while a flash of the
intensive irradiation is observed on the PC stop-zone boundary near
$\lambda = 562.5$ nm. This emission is shifted to the
long-wavelength region in comparison with the spectral position of
the absorption of the initial radiation by the stop-zone, and the
emission intensity was $\approx 0.01$ of that of the excitation
radiation. Changes in the spectra of synthetic opals occur also when
the crystal thickness grows \cite{7}.

In \cite{2}, it was shown that photonic crystals display the effect
of ``hidden box'' for biological molecules infiltrated into the opal
pores. Spherical gold nanoparticles in a colloidal solution show a
plasmon resonance close to the opal stop-zone. In the present paper,
the Bragg reflectance and the secondary emission (photoluminescence)
of synthetic opals infiltrated by aqueous colloidal gold, glycine,
and a complex of colloidal gold with glycine are studied. A possible
mechanism of the observed effects is discussed.

\begin{figure*}
\includegraphics[width=8.4cm]{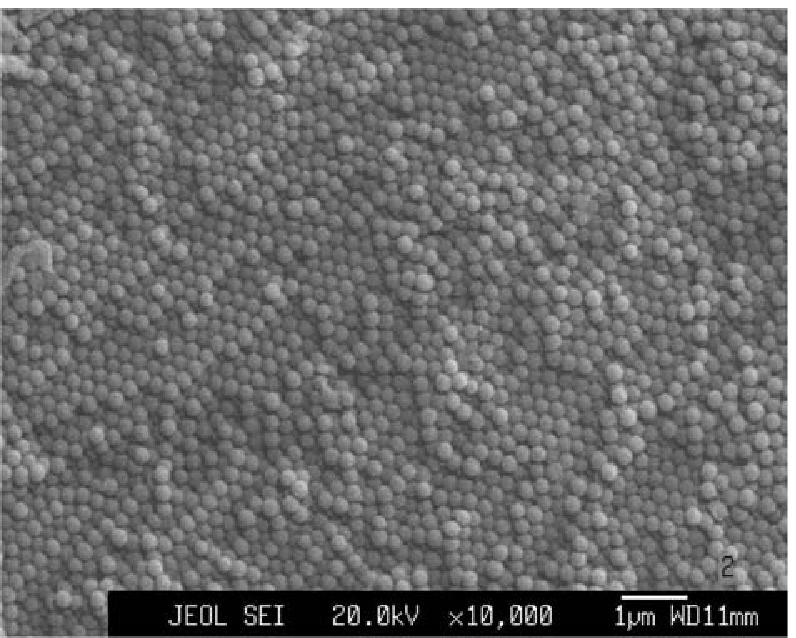}\hspace{0.5cm}\includegraphics[width=8.5cm]{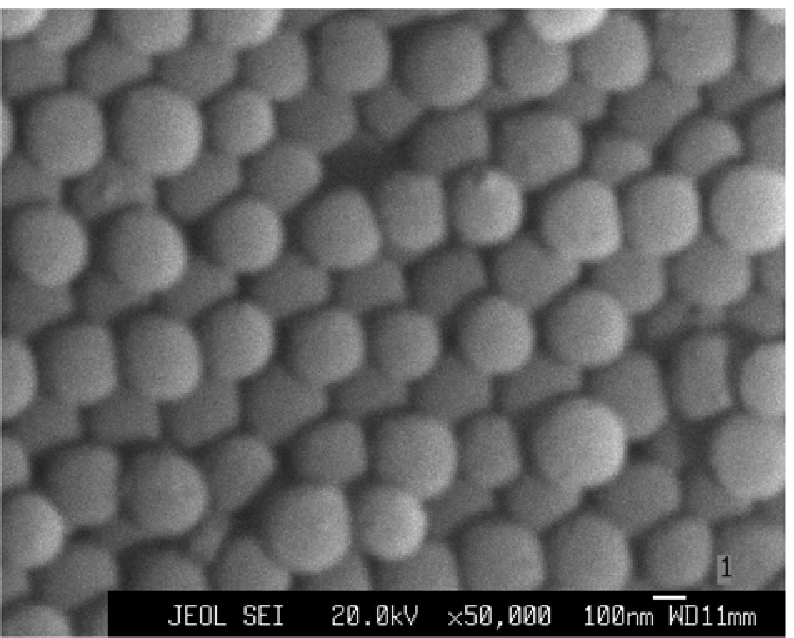}\\
{\Large\it a\hspace{8.4cm}b} \vskip-3mm\caption{Microstructure of
synthetic opal with magnifications $k = 10000$ ({\it a}) and $k =
50000$ ({\it b})}
\end{figure*}

\section{Materials and Methods}\vspace*{-1.5mm}

Nanodisperse silica globules were synthesized by the Stober method,
followed by the natural sedimentation and the annealing in air at a
temperature of $600~^{\circ}$C. The annealing is necessary to remove
the organic components of a buffer solution from the pores of opal
\cite{1}. The size of the fabricated opal crystals was $10\times 10
\times 2$ mm. For the infiltration of PC, we have used a colloidal
aqueous solution of 10--20-nm gold nanoparticles (40 mg/l), a
1-mg/ml aqueous solution of an $\alpha$-Gly powder (Sigma), as well
as a complex of colloidal gold with glycine molecules prepared as a
result of mixing the above-mentioned solutions according to the
ratio: 5 volume parts of the aqueous glycine solution, 2 -- the
colloidal aqueous gold solution, and 3 -- distilled water. The
process of infiltration has been made by the multiple ``drop and
dry'' procedure of the corresponding solution (10 mcl) on the PC
surface and the consecutive drying of them at room temperature. The
structure of the samples and optical properties have been
characterized with SEM analysis, optical spectroscopy of the visible
range, and luminescence. SEM images of opals were obtained with an
EPMA SEI JXA-8200 microscope. The reflectance of nonpolarized light
in the visible spectral range was measured with a spectrophotometer
based on a DMR-4 monochromator at different (10$^{\circ}$,
20$^{\circ},$ and 30$^{\circ}$) angles of incidence. The spectra of
fluorescence have been registered with a Perkin Elmer LS-55
fluorescence spectrometer under excitation with $\lambda_{\rm exc}$
= 255 and 370 nm with long-pass filters, 290 and 390 nm,
respectively.

\section{Results of Experiments}\vspace*{-1.5mm}

The SEM images of synthetic opal are presented in Fig.~1. The
diameter ($d$) of globules was estimated as 240 nm, and the size of
cavities as 30--50~nm.

The reflection spectra of initial synthetic opals, opals with
colloidal gold, and opal with complex of colloidal gold and glycine
in the visible region are presented in Fig. 2. The Bragg reflection
for initial opal is observed in the region of 470--510 nm. This
region displays the stop-zone of PC. The obtained data correlate
well with the calculation of (111) Bragg maximum for a photonic
crystal built from silica globules with the diameter $d = 240$~nm
according to the formula
\begin{equation}
\lambda_{\rm B}=2\sqrt{\frac{2}{3}} d\sqrt{n^{2}_{\rm
eff}-\sin^{2}\varphi}.
\end{equation}

Here, the effective refractive index $n^{2}_{\rm eff}=n^{2}_{{\rm
SiO}_{2}}$ $ {f_{{\rm SiO}_{2}} + n^{2}_{m} (1-f_{{\rm SiO}_{2}})}$,
where $n_{{\rm SiO}_{2}} = 1.47$, $n_{m} = 1$ for air or $n_{m} =
1.33$ for the aqueous solution, $f = 0.74$ is the filling factor for a dense
packing of equal spheres. As one can see from Fig.~2,({\it a--c}),
the Bragg reflection maximum displays a short-wavelength shift from
505 nm for initial opal to 500 nm after the introduction of colloidal
gold in PC and to 495 nm after the introduction of a colloidal
gold-glycine complex. Additionally, after the introduction of colloidal
gold and colloidal gold with glycine in PC, an increase
of the Bragg maximum intensity by 1.5 and 3 times, respectively, is observed.

\begin{figure}
\includegraphics[width=6.5cm]{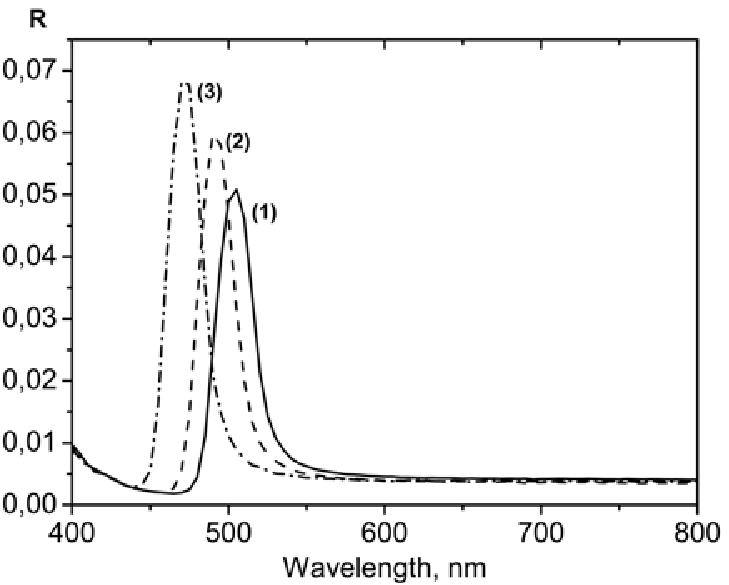}\\
{\Large\it a}\\ [2mm]
\includegraphics[width=6.5cm]{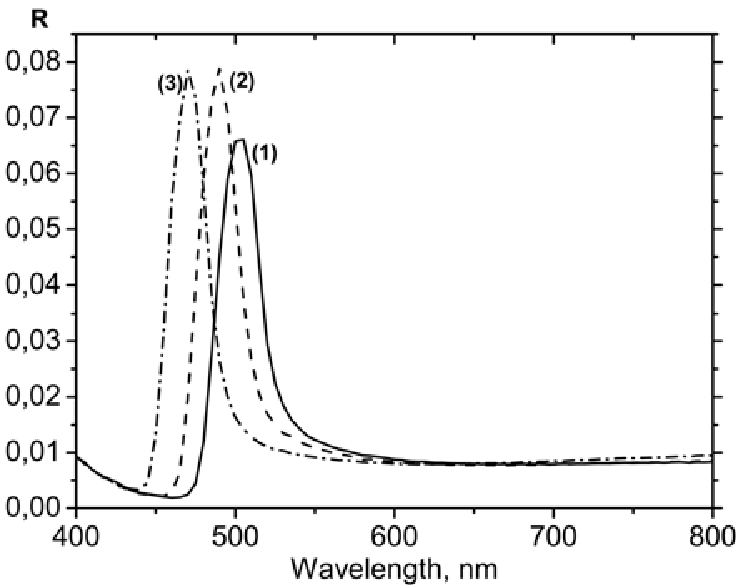}\\
{\Large\it b}\\ [2mm]
\includegraphics[width=6.5cm]{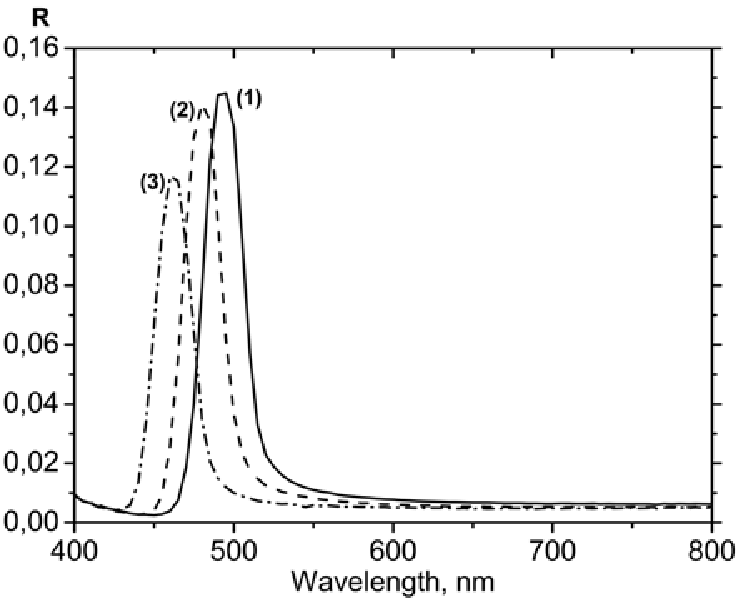}\\
{\Large\it c} \vskip-3mm\caption{Spectra of the Bragg reflection of
{\it a}-- initial opal, {\it b} -- opal with colloidal gold, {\it c}
-- opal with colloidal gold and glycine, measured at the incidence angles of
$10^{\circ}$ (curves {\it 1}), $20^{\circ}$ (curves {\it
2}), and $30^{\circ}$ (curves {\it 3})}
\end{figure}

We have registered the photoluminescence excitation spectra at
$\lambda = 500$ nm at the normal incidence of light to the surface of PC in
the backscattering geometry. The excitation spectra show 3 intense
bands (Fig.~3) with maxima at 236, 250, and 360~nm. In \cite{4}, the
band at 250 nm was attributed to the zone-zone transition in SiO$_2$.
Under the infiltration of PC with colloidal gold and glycine, the band
at 360 nm was shifted to the long wavelength region.

\begin{figure}
\includegraphics[width=6.3cm]{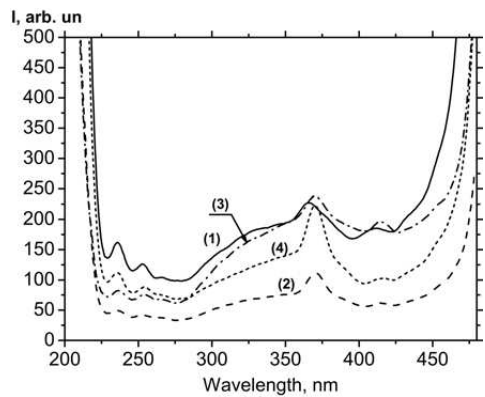}
\vskip-3mm\caption{Excitations spectra of fluorescence (registered
at $\lambda = 500$ nm) for initial PC (curve {\it 1}), PC with
colloidal gold (curve {\it 2}), PC with glycine (curve {\it 3}), PC
with colloidal gold and glycine (curve~{\it 4})}
\end{figure}

\begin{figure}[h]
\includegraphics[width=6.5cm]{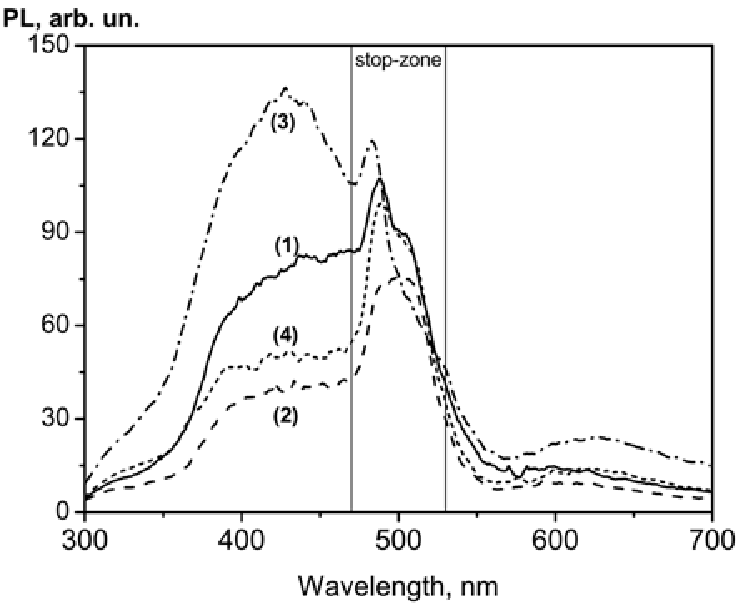}\\
{\Large\it a}\\ [2mm]
\includegraphics[width=6.5cm]{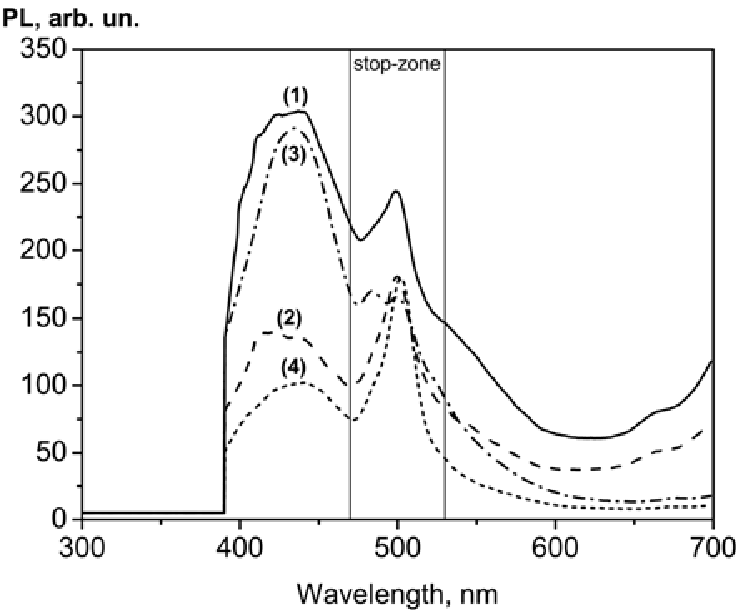}\\
{\Large\it b}\\
\vskip-3mm\caption{Photoluminescence spectra: {\it 1} -- initial PC;
{\it 2} -- PC with colloidal gold; {\it 3} -- PC with glycine; {\it
4} -- PC with colloidal gold and glycine. $a$ -- excited by
$\lambda_{\rm exc}$ = 255 nm and $b$ -- excited by $\lambda_{\rm
exc}$ = 370 nm}
\end{figure}

The photoluminescence spectra have been registered for initial opal,
as well as for opals infiltrated by colloidal gold, glycine, and
complex of colloidal gold with glycine  (Fig. 4). After the
infiltration of colloidal gold, a decrease of the intensity of the
band at 375--450 nm was observed. Under the introduction of glycine
in opal, the intensity of the peak near 500 nm is suppressed. The
observed features in the photoluminescence spectra could be
explained by the close location of the PC stop-zone and the plasmon
resonance region in gold nanoparticles to the region of emission of
initial opal and the fluorescence property of gold particles
\cite{8}. Thus, it was shown in \cite{9} that gold particles have a
plasmon resonance near 530 nm. Meanwhile, the gold particles or gold
rough surface show the photoluminescence in the visible region
\cite{8}. This effect could lead to the formation of a complex
fluorescence band of PC with colloidal gold and suppress an
influence of the stop-zone in the region of appearance of plasmonic
effects. However, at a certain frequency, we could get an essential
enhancement of the emission signal as well.\looseness=1

Three bands become apparent in the spectra of secondary emission of
initial and modified synthetic opal under the excitation with
$\lambda = 255$ (Fig. 4,{\it a}) and 370 nm (Fig. 4,{\it b}): 1)
375--450 nm, 2) 490-500 nm, 3) 650-670 nm. These bands are connected
with defects and admixtures. For example, according to \cite{10},
the band at 523 nm could be assigned to the surface state of
$\equiv$Si--H with an energy of 2.37 eV; the band at 625 nm could be
assigned to the volume state of $\equiv$Si--O with an energy of 1.9
eV, and the band at 692 nm -- to the surface state of $\equiv$Si--O
(1.79 eV). However, the band in the region near 400 nm is under
discussions \cite{10}, and it can be connected with different
admixtures in chemically grown SiO$_2$ globules, for example, with
ZrO$_2$ \cite{11}, {\it etc}.\looseness=1

Due to the infiltration of PC with colloidal gold, we have observed
the enhancement of the emission at 500 nm by 1.5 times and the
suppression of the emission at 375--450 nm (Fig. 4,{\it a}). The
infiltration of opal by glycine leads to the enhancement of a wide
emission band near 434 nm by 3 times with a simultaneous decrease of
the band at 500 nm. In the case of the infiltration of opal with the
complex of colloidal gold with glycine, we have registered an
increase of the intensity of both 500 nm and 434 nm bands by 2 times
simultaneously. These changes could be caused by the enhancement of
a local field near the surface of PC, as well as by plasmonic
effects due to the presence of colloidal gold. Note that colloidal
gold in an aqueous solution under the excitation with $\lambda =
255$ and 370 nm reveals a very weak emission in the region of
420--450 nm (Fig. 5). The intensity of this emission is about two
orders less than the emission from opal.\looseness=1

\begin{figure}
\includegraphics[width=6.3cm]{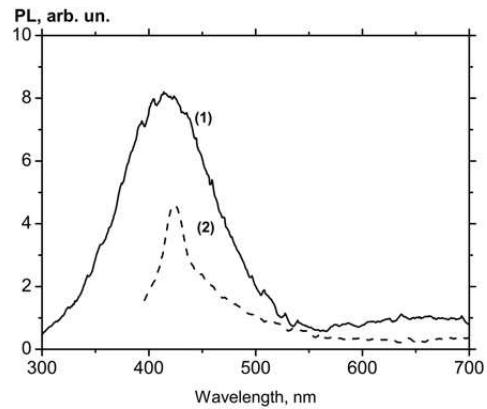}
\vskip-3mm\caption{Photoluminescence spectra of 10--20-nm colloidal
gold in an aqueous solution excited with $\lambda_{\rm exc} = 255$
(curve {\it 1}) and $\lambda_{\rm exc} = 370$ nm (curve {\it 2})}
\end{figure}

\section{Conclusions}

The investigation of the Bragg reflection and secondary emission
spectra of synthetic opals, opals infiltrated by colloidal gold,
glycine, and the complex of colloidal gold with glycine is
performed. The Bragg reflection band of PC is shifted by 5--15~nm to
the short wavelength side under the infiltration of opal, and the
intensity of this band increases by 1.5-3 times. Opal reveals a
complex photoluminescence band in a vicinity of 350--550~nm. The
shape of the emission band connected with defects in synthetic opal
is determined by the excitation wavelength and the type of
infiltrated substances. The suppression of the emission band
(375--450~nm) and an enhancement of the shoulder (470--510~nm) of
the stop-zone band  under the infiltration of opal with colloidal
gold and colloidal gold with glycine could be caused by plasmonic
effects and the influence of the increased density of photonic
states on the boundary of the PC stop-zone.

\vskip3mm We thank Ukrainian-Russian project 4/11-24 ``The glow of
three-dimensional photonic crystals for optical and electrical
excitation'' for the financial support.

\rezume{%
ВТОРИННА ЕМІСІЯ СИНТЕТИЧНИХ ОПАЛІВ,\\ ІНФІЛЬТРОВАНИХ КОЛОЇДНИМ
ЗОЛОТОМ\\ ТА ГЛІЦИНОМ}{Г.І. Довбешко, O.M. Фесенко, В.В. Бойко, В.Р.
Романюк,\\ В.С. Горєлік, В.М. Моісеєнко, В.Б. Соболєв, В.В.
Швалагін}{Проведено порівняльний аналіз вторинної емісії
(фотолюмінесценції) та брегівського відбивання фотонних кристалів
(синтетичних опалів), інфільтрованих колоїдним золотом, гліцином та
комплексом колоїдного золота з гліцином. Інфільтрація колоїдного
золота та його комплексу з гліцином в пори фотонного кристала
привела до короткохвильового (на 5--15 нм) зсуву максимуму
брегівського відбиття та зростання його інтенсивності в 1,5--3 рази.
У фотолюмінесценції інфільтрація колоїдного золота та комплексу
колоїдного золота з гліцином у пори фотонного кристала приводить до
пригнічення смуги поблизу 375--450 нм та до підсилення смуги поблизу
краю стоп-зони на 470--510 нм. Форма смуги фотолюмінесценції опалу,
що викликана його дефектами та домішками, визначається довжиною
хвилі збуджуючого випромінювання та типом інфільтрату. Обговорено
можливі механізми ефектів, що спостерігаються.}

\end{document}